\documentclass[a4paper,11pt]{article}
\pdfoutput=1 

\usepackage{jinstpub} 

\title{\boldmath A simulation tool for MRPC telescopes of the EEE project}

\author[1,2,3,*]{G.~Mandaglio,\note[*]{Corresponding author.}} 
\author[4,5]{M.~Abbrescia,}
\author[1,6]{C.~Avanzini,}
\author[1,6,7]{ L.~Baldini,}
\author[1,8]{ R.~Baldini~Ferroli,}
\author[1,6,7]{G.~Batignani,}
\author[1,9,10]{ M.~Battaglieri,}
\author[1,11,12]{ S.~Boi,}
\author[13]{E.~Bossin,}
\author[1,14,15]{ F.~Carnesecchi,}
\author[1,12]{ C.~Cical\`{o},}
\author[1,14,15]{L.~Cifarelli,}
\author[1]{ F.~Coccetti,}
\author[1,16]{ E.~Coccia,}
\author[1,17]{A.~Corvaglia,}
\author[18,19]{ D.~De~Gruttola,} 
\author[18,19]{S.~De~Pasquale,}
\author[1,8]{F.~Fabbri,}
\author[2]{ A. Fulci,}
\author[15]{D.~Falchieri,}
\author[1,20,21]{ L.~Galante,}
\author[1,15]{M.~Garbini,}
\author[10]{ G.~Gemme,}
\author[1,22]{ I.~Gnesi,}
\author[1]{S.~Grazzi,}
 \author[1,13,15]{D.~Hatzifotiadou,}
 \author[1,3,23]{ P.~La~Rocca,}
 \author[24]{ Z.~Liu,}
 \author[25]{ L.~Lombardo,}
 \author[26]{G.~Maron,}
 \author[5]{ M.~N.~Mazziotta,}
\author[11,12]{A.~Mulliri,}
\author[1,15]{ R.~Nania,}
\author[1,15]{ F.~Noferini,}
\author[27]{F.~Nozzoli,}
\author[1,14]{ F.~Palmonari,}
\author[17,28]{ M.~Panareo,}
\author[1,17]{M.~P.~Panetta,}
\author[6,29]{ R.~Paoletti,}
 \author[25]{ M.~Parvis,}
\author[1,26]{ C.~Pellegrino,}
\author[1,10]{L.~Perasso,}
\author[1,15]{O.~Pinazza,}
\author[1,3,23]{ C.~Pinto,}
\author[1,8]{S.~Pisano,}
\author[1,3,23]{ F.~Riggi,}
\author[1]{ G.~Righini,}
\author[18,19]{C.~Ripoli,}
\author[5]{ M.~Rizzi,}
\author[1,14,15]{ G.~Sartorelli,}
\author[1,15]{E.~Scapparone,}
\author[22,30]{ M.~Schioppa,}
\author[29]{ A.~Scribano,}
\author[1,15]{M.~Selvi,}
\author[1,11,12]{ G.~Serri,}
\author[10,31]{ S~ Squarcia,}
\author[10,31]{M.~Taiuti,}
\author[1,6]{ G.~Terreni,}
\author[1,2,3]{ A.~Trifir\`{o},}
\author[1,2,3]{M.~Trimarchi,}
\author[2]{ A.S. Triolo,} 
\author[26]{C.~Vistoli,}
\author[16]{ L.~Votano,}
\author[1]{M.~C.~S.~Williams,}
\author[1,14,15]{ A.~Zichichi,}
\author[1]{ R.~Zuyeuski}

\affiliation[1]{Museo Storico della Fisica e Centro Studi e Ricerche "Enrico Fermi", Roma, Italy}

\affiliation[2]{Dipartimento di Scienze Matematiche e Informatiche, Scienze Fisiche e Scienze della Terra, Universit\`{a} di Messina, Messina, Italy}

\affiliation[3]{INFN Sezione di Catania, Catania, Italy }

\affiliation[4]{Dipartimento Interateneo di Fisica, Universit\`{a} di Bari, Bari, Italy }

\affiliation[5]{INFN Sezione di Bari, Bari, Italy}

\affiliation[6]{INFN Sezione di Pisa, Pisa, Italy}

\affiliation[7]{Dipartimento di Fisica, Universit\`{a} di Pisa, Pisa, Italy }

\affiliation[8]{INFN Gruppo Collegato di Cosenza, Laboratori Nazionali di Frascati (RM), Italy}

\affiliation[9]{Thomas Jefferson National Accelerator Facility, Newport News, VA 23606, USA }

\affiliation[10]{INFN Sezione di Genova, Genova, Italy}

\affiliation[11]{Dipartimento di Fisica, Universit\`{a} di Cagliari, Cagliari, Italy}

\affiliation[12]{INFN Sezione di Cagliari, Cagliari, Italy }

\affiliation[13]{CERN, Geneva, Switzerland }

\affiliation[14]{Dipartimento di Fisica, Universit\`{a} di Bologna, Bologna, Italy}

\affiliation[15]{INFN Sezione di Bologna, Bologna, Italy }

\affiliation[16]{Gran Sasso Science Institute, Italy}

\affiliation[17]{INFN Sezione di Lecce, Lecce, Italy}

\affiliation[18]{Dipartimento di Fisica, Universit\`{a} di Salerno, Salerno, Italy}

\affiliation[19]{INFN Gruppo Collegato di Salerno, Salerno, Italy}

\affiliation[20]{Dipartimento di Scienze Applicate e Tecnologia, Politecnico di Torino, Torino, Italy}

\affiliation[21]{INFN Sezione di Torino, Torino, Italy }

\affiliation[22]{INFN Gruppo Collegato di Cosenza, Laboratori Nazionali di Frascati (RM), Italy }
\affiliation[23]{Dipartimento di Fisica e Astronomia, Universit\`{a} di Catania, Catania, Italy }
\affiliation[24]{ICSC World laboratory, Geneva, Switzerland }
\affiliation[25]{Dipartimento di Elettronica e Telecomunicazioni, Politecnico di Torino, Torino, Italy}
\affiliation[26]{INFN-CNAF, Bologna, Italy }
\affiliation[27]{INFN Trento Institute for Fundamental Physics and Applications, Trento, Italy}
\affiliation[28]{Dipartimento di Matematica e Fisica, Universit\`{a} del Salento, Lecce, Italy}
\affiliation[29]{Dipartimento di Scienze Fisiche, della Terra e dell'Ambiente, Universit\`{a} di Siena, Siena, Italy}
\affiliation[30]{Dipartimento di Fisica, Universit\`{a} della Calabria, Rende (CS), Italy }
\affiliation[31]{Dipartimento di Fisica, Universit\`{a} di Genova, Genova, Italy}
\emailAdd{gmandaglio@unime.it}

\abstract{The Extreme Energy Events (EEE) Project is mainly devoted to the study of the secondary cosmic ray radiation by using muon tracker telescopes made of three Multigap Resistive Plate Chambers (MRPC) each. The experiment consists of a telescope network  mainly distributed across Italy, hosted in different building structures pertaining to high schools, universities and research centers. Therefore, the possibility to take into account the effects of these structures on collected data is important for the large physics programme of the project.
A simulation tool, based on GEANT4 and using GEMC framework, has been implemented to take into account the muon interaction with EEE telescopes and to estimate the effects on data of the structures surrounding the experimental apparata.A dedicated event generator producing realistic muon distributions, detailed geometry and microscopic behavior of MRPCs have been included to produce experimental-like data. The comparison between simulated and experimental data, and the estimation of detector resolutions  is here presented and discussed.}


\arxivnumber{1234.56789} 

\collaboration[c]{}

\proceeding{XV Workshop on Resistive Plate Chambers and Related Detectors RPC2020\\
  10-14 February, 2020\\
  Rome (Italy)}

\begin{document}
\maketitle
\flushbottom

\section{Introduction}
\label{sec:intro}
The EEE experiment\cite{Ref1} is a project of the   \emph{``Museo Storico della Fisica e Centro Studi e
Ricerche Enrico Fermi''}\cite{Ref2} in collaboration with  \textit{``Istituto Nazionale di Fisica Nucleare''} (INFN) \cite{Ref3},  and \emph{"Ministero dell'Universit\`{a},
dell'Istruzione e della Ricerca"} (MIUR) and CERN\cite{Ref4}.
The experiment consists of a network of Multigap Resistive Plate Chambers (MRPC) based telescopes, located mainly in Italian High Schools,  at CERN and in some INFN sections, spanning a total area of about 0.3$\times$10$^6$ km$^2$.
The experiment, during 5 runs of data taking, collected 
more than 100 billion  of candidate muon tracks
allowing to pursue a varied scientific programme:  extensive air shower investigation via coincidence between different telescope \cite{riggilong,Ref14}, investigation of Forbush decrease\cite{forby}, monitoring of long-term stability of civil structures \cite{riggistab} etc.
To fully understand the data, a precise knowledge of the effect on the measurements of the structures surrounding and hosting the detectors is needed.
For these reasons, a simulation tool, implemented using the GEMC (GEant4 MonteCarlo) framework was developed, in order  to estimate the telescope absolute efficiency and angular acceptance, and, starting from them, the single muon rates.

In the present paper, the simulation tool is described and some interesting results   obtained with the simulated data, like the reproduction of the experimental condition of a telescope working in a laboratory with a peculiar structure and position of the building, the estimation of the detector efficiency and the effect of the material surrounding the telescope on the ability of the detector to measure the muon arriving direction are shown. A comparison between  the experimental and simulated polar angle distributions is also presented and discussed.

\section{The EEE Detectors}

An EEE telescope consists of three MRPCs
with a 80$\times$160 cm$^2$ 
active area each, 
assembled - in the most common configuration - in a  stack with 50 cm distance between the chambers.
Readout panel of each chamber are  segmented in  24 copper strips (180 cm $\times$ 2.5 cm spaced by 7 mm), 
which  collect the signals produced in the gas --mixture of C$_2$F$_4$H$_2$ (98\%) and SF$_6$ (2\%)-- contained in the chambers by the impinging charged particles.
The chamber readout provides two coordinates for each hit:  one is given by the coordinate of the fired strip or, in the case of contiguous strips fired, by their average value, while the other is obtained by  the time difference of the signals measured at the opposite edges of the strip. For details on the detector see Ref. \cite{degruttola} and references therein.

\section{ The Simulation Tool}

\begin{figure}[h]
\centering
\includegraphics[width=0.39\columnwidth]{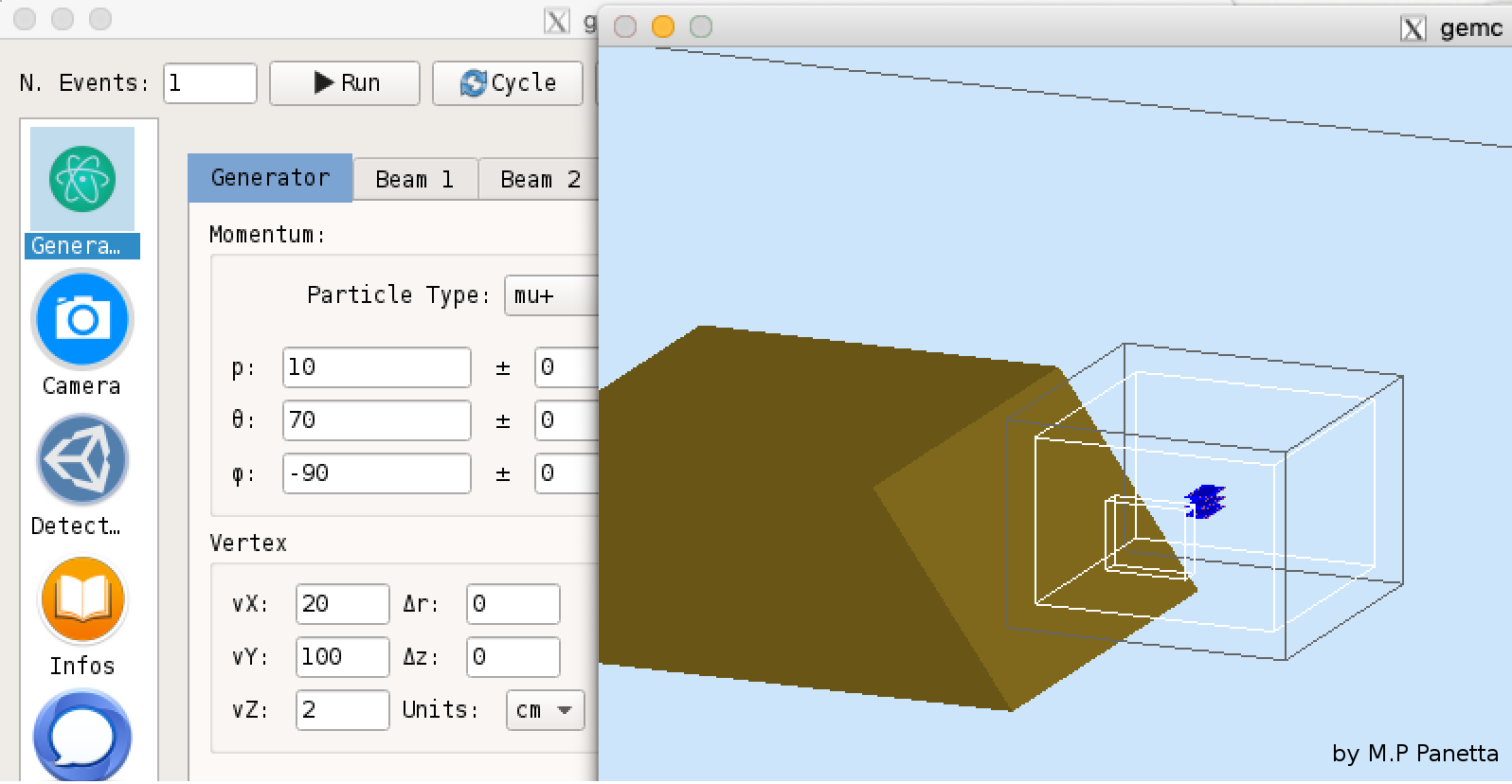}\hspace{0.5cm}
\includegraphics[width=0.2\columnwidth]{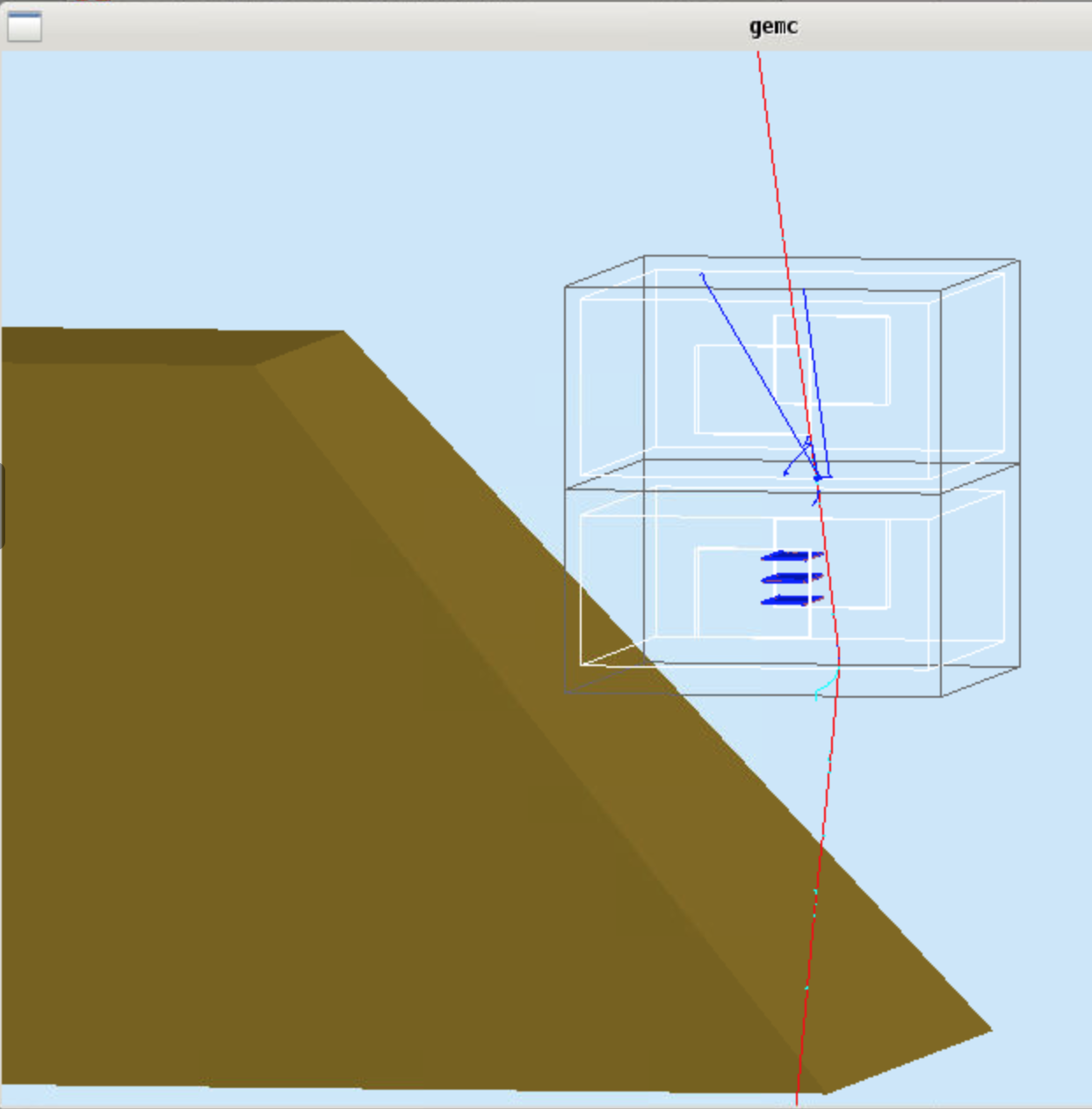}\hspace{0.5cm}
\includegraphics[width=0.27\columnwidth]{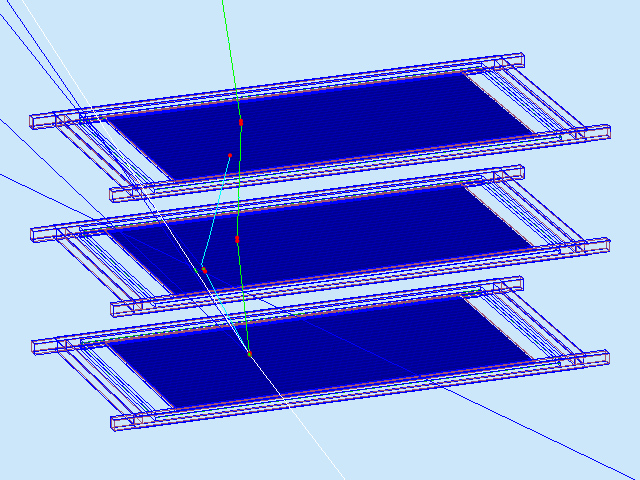}
\caption{Left panel: GEMC graphical interface. Central panel: details of a muon interaction  with the floor of two rooms (in this case, 30 cm thick  concrete) and the detector. Right panel: a detail of the telescope.} \label{geo}
\end{figure}

The simulation tool used for modeling the EEE telescopes is based on  the GEMC \cite{gemc}   framework which provides a simple way to build up a  user-defined geometry and hit description. 
Detector and building structures are implemented  using  the standard GEANT volume description. The program handles multiple input/output format and provides a graphical interface to visualize the detector and the hits in active and passive volumes (see figure \ref{geo}).

GEMC supports the use of an external event generator, with data written  in the Lund format, and it is provided with an internal event generator based on the  model described in Ref.\cite{egpar} used in this work to generate the single-muon distribution.
The used parametrization is able to well reproduce the existing measurements \cite{egpar}.  
The absolute muon flux normalization, used in this simulation, is the one reported in the PDG \cite{pdg}.

MRPC response was parametrized on the base of the measured performance of the chambers \cite{degruttola}. In particular, the algorithm  mimicking the avalanche propagation in the gas is effectively described by a cone with vertex generated  at the interaction point in the chamber and developing toward the anode (in this case positioned downwards). 
The room hosting the detector is parameterized by a possible box of concrete. Of course, more complicated geometries are customizable too, as one can see in the left and central panels of Figure \ref{geo}, where the room hosting the EEE telescope is located in a building on a steep mountain side, as it really is for one of the EEE telescopes.

The information generated by GEMC and necessary to reconstruct the muon track is: the total number of hits for each chamber;  the coordinates of the strips giving signals;  the signal time from the generation point to the edges of the chamber. By using this information the simulation program is able to write output data files which are then fed to the usual EEE reconstruction chain. The  reconstruction code efficiency is found higher than 99\%.


\section{Simulation results}
In this section a study on the validation of the simulation, performed by comparing the simulated polar angle distribution corrected by the experimental efficiency to the experimental one is reported. Moreover, two investigations about the effects  of the material surrounding the telescope on the collected data are discussed.

\subsection{Comparison between Experimental and Simulated Data}

In order to compare   simulated and experimental data, the detector efficiency has to be carefully estimated. Therefore,  a telescope was selected for its stable working condition and negligible shielding of the hosting room (telescope TORI-03, located in a high school in Turin) to calculate the efficiency and to compare the simulated polar angle distribution to the experimental one. 
The efficiency of the telescope is performed by dividing each chamber into  24$\times$20 $\rm cm^2$ sectors, and then by estimating for each sector  the tracking efficiency and the counting efficiency, assuming that no correlation exists between the two quantities. 


\begin{figure}[h]
\centering
\includegraphics[width=0.45\columnwidth]{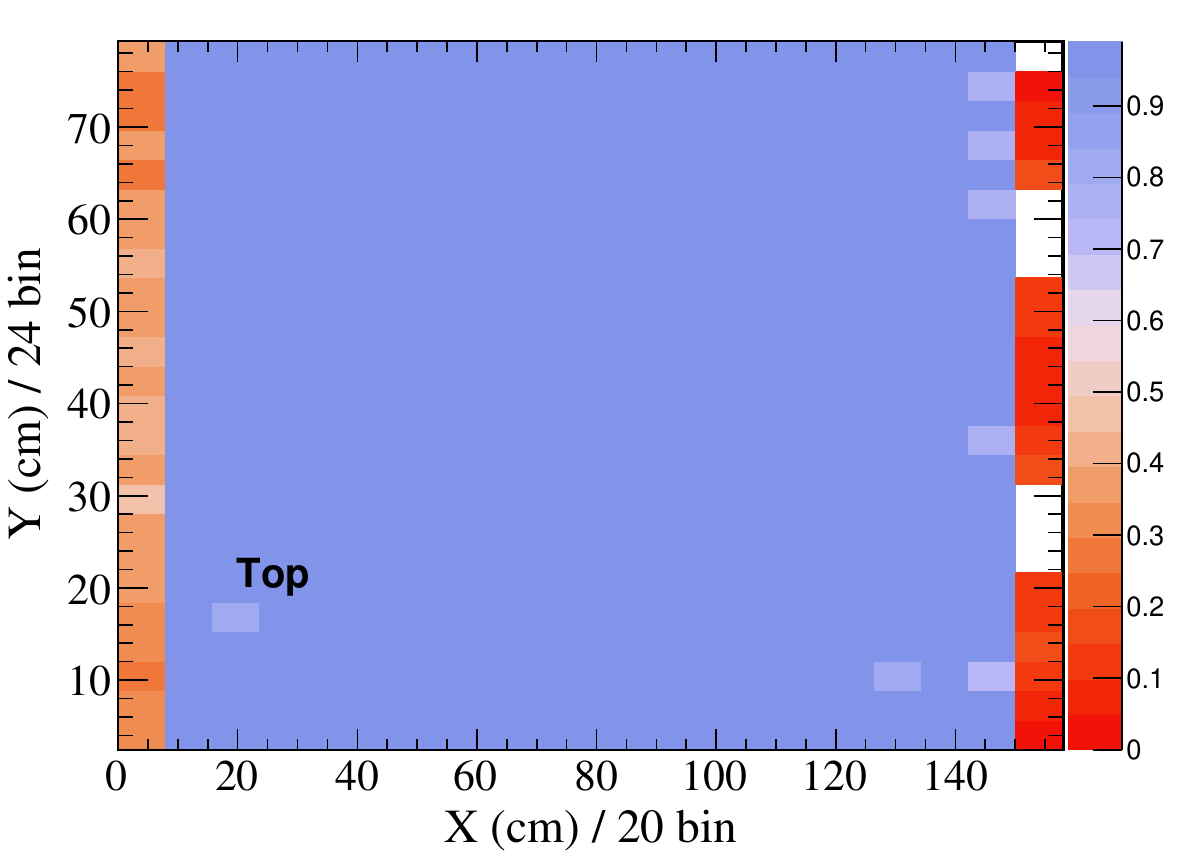}
\put(-100,90){Preliminary}\hspace{0.3cm}
\includegraphics[width=0.45\columnwidth]{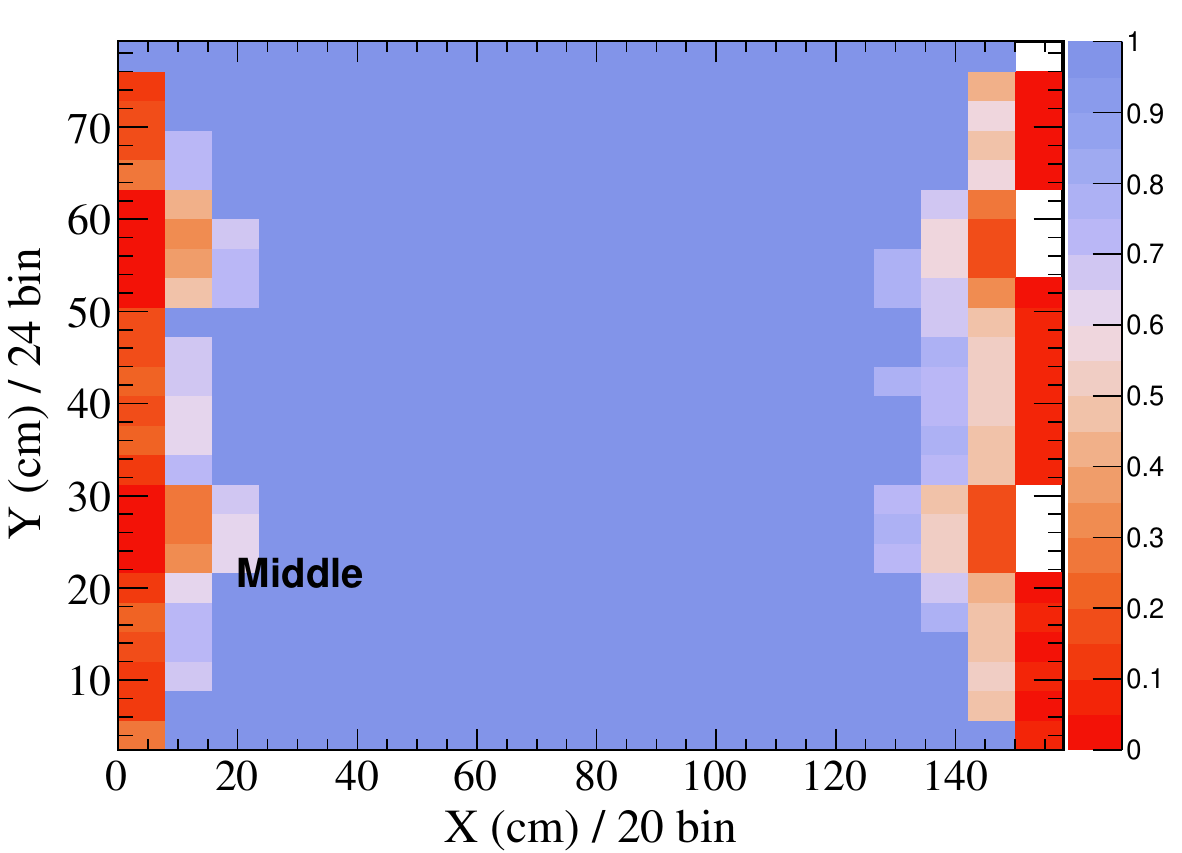}
\put(-100,90){Preliminary}\\
\includegraphics[width=0.45\columnwidth]{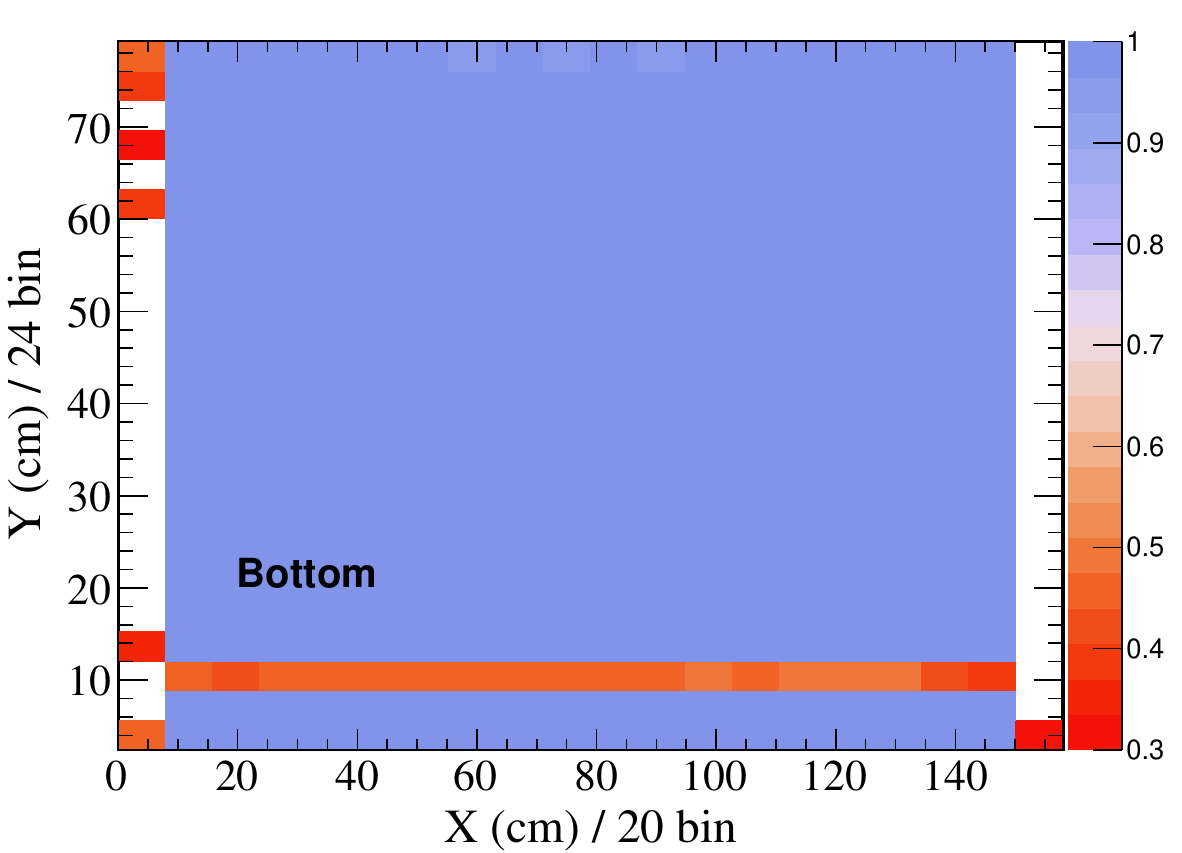}
\put(-100,90){Preliminary}\hspace{0.3cm}
\includegraphics[width=0.45\columnwidth]{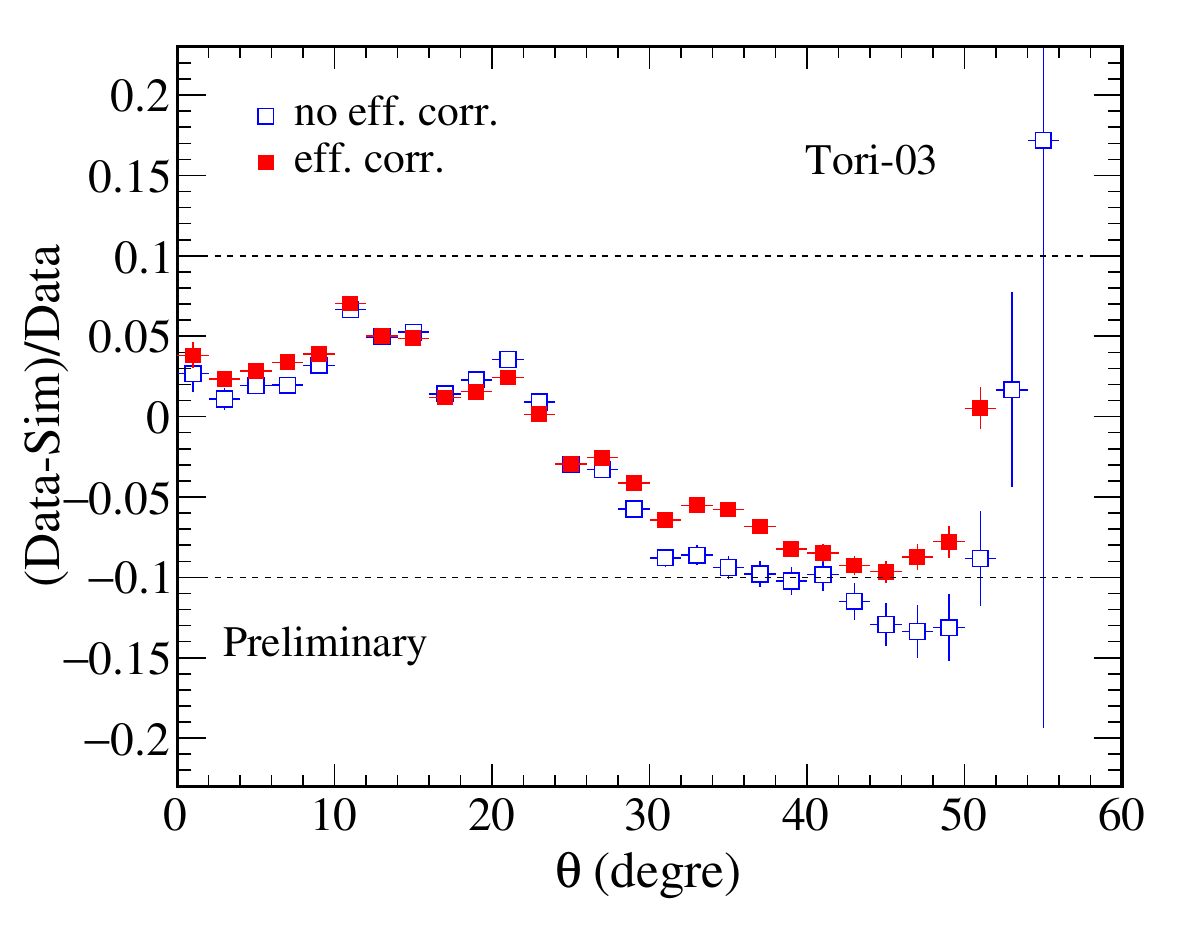}
\caption{Global efficiency map for top, middle and bottom chamber of TORI-03 telescope (top-left, top-right and bottom-left respectively. Bottom-right panel the experimental-simulation data ratio normalized to experimental data of the polar angle distribution, without (empty squares) and with (full squares)  efficiency correction.} \label{effi}
\end{figure}



Tracking efficiency is determined starting from experimental data, using the hits from two chambers, fitting a track to the hits, and extrapolating it on the third. If a hit is found inside the 2D histogram bin where the predicted intersection point is located, the hit is classified as ``detected'' and the relative map is filled. If not, the hit is classified as ``missing'' and another map is filled. The tracking efficiency map is defined as the ratio between the map of detected hits and the map of missing ones.

   Counting efficiency is determined from experimental data as well, starting from a map from each chamber, filled, without applying any cut, with the hits on that chambers and correcting it for the effects due to the geometrical acceptance and then normalizing it.


The global efficiency map, for each chamber, is  obtained as the product of the tracking and counting efficiency maps of the same chamber.
More details about the procedure to measure the detector efficiency are reported in Ref. \cite{mandy}.

 The global efficiencies of the three chambers of  the TORI-03 telescope are shown in the top panels and bottom-left panel of figure~\ref{effi};  these maps are used to correct the polar angle distribution of the simulated events.
 These efficiency maps are also useful to locate inefficiency in small regions in a chamber (see top-left panel of Fig. \ref{effi}), or at a border of the chamber where a reduced gas flow is likely (see top-right panel of Fig. \ref{effi}) or corresponding to a whole strip (see bottom-left panel of Fig. \ref{effi}). 
 
In the bottom-right panel of figure \ref{effi}, the difference in the angular distributions obtained with experimental and simulated data, normalized to the experimental one, with and without efficiency correction is reported. The efficiency corrections derived from the data allow to improve the agreement between experimental and simulation distributions within 10\% in the whole polar angular acceptance of the  EEE telescope.       
The improvement of the experimental-simulation agreement at large angle proves the procedure reliability in the estimation of the telescope efficiency  by using the experimental data. However, further investigations on this promising method to directly estimate the efficiency from data are in progress.

\subsection{Muon absorption by the structures surrounding the telescopes}

\begin{figure}[h]
\centering
\includegraphics[width=0.6\columnwidth]{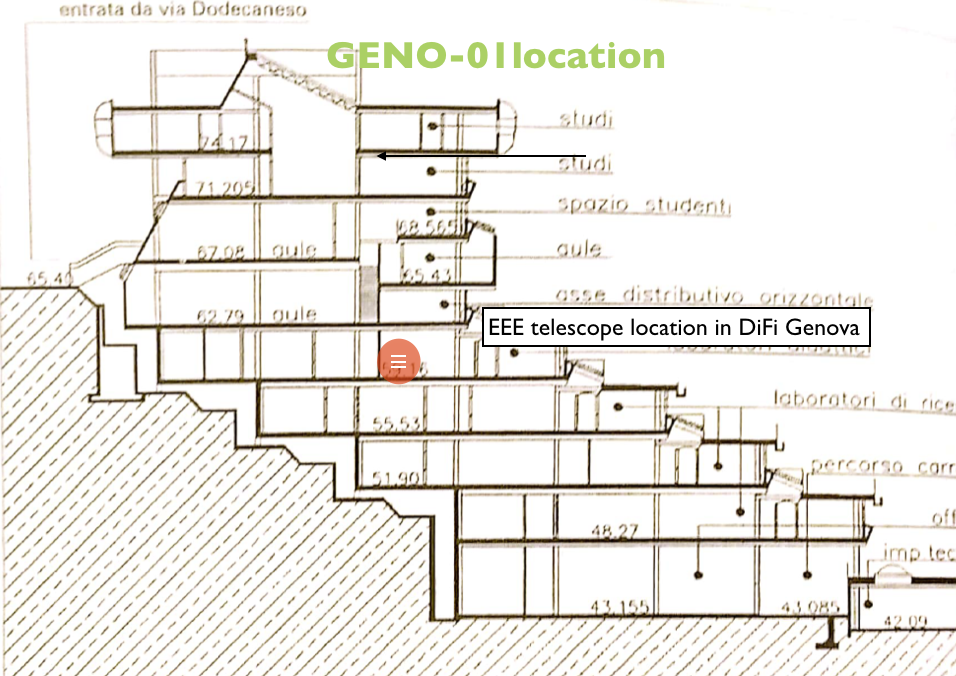}
\caption{Scheme of the Department of Physics of the University of Genoa.\label{gen}}
\end{figure}
\begin{figure}[h]
\centering
\includegraphics[width=0.5\columnwidth]{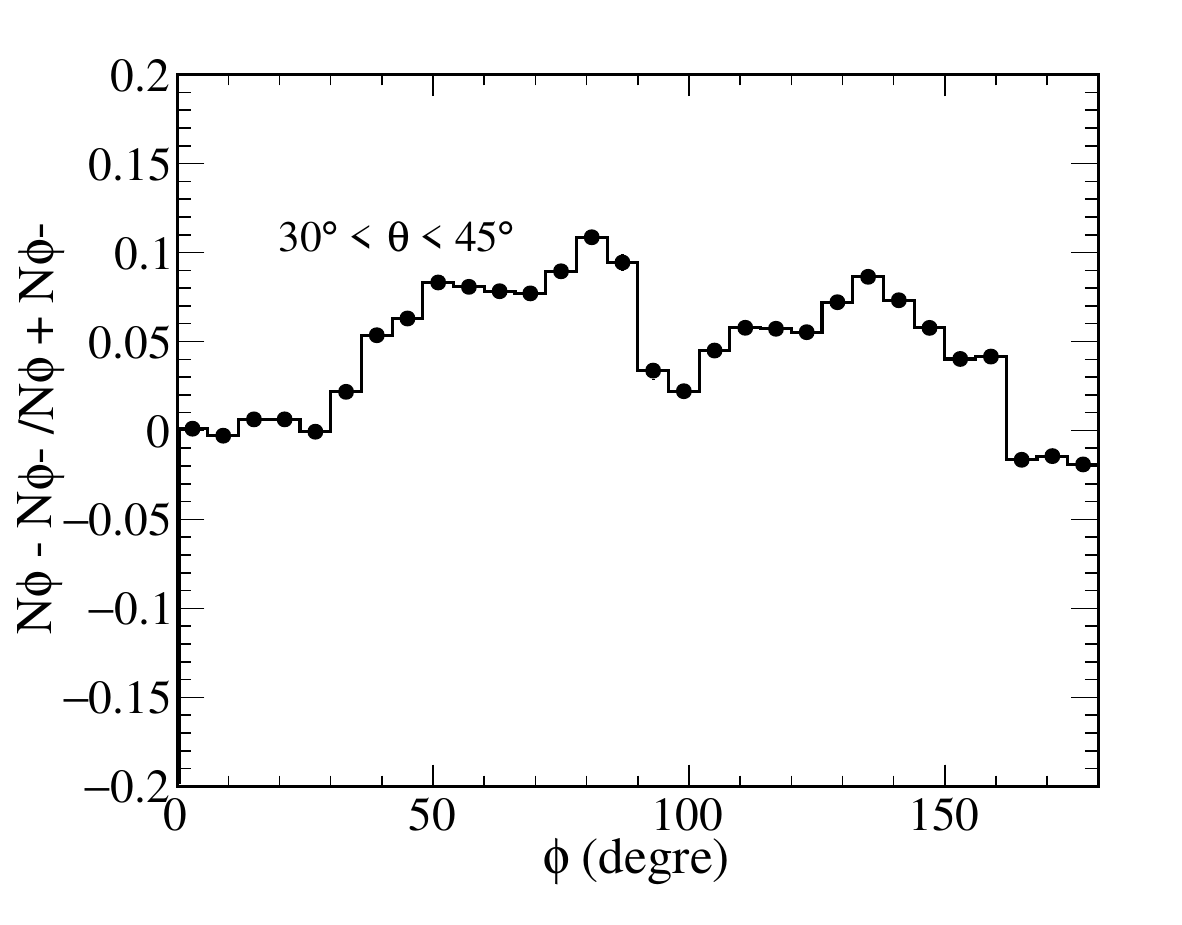}
\put(-130,50){Preliminary}
\includegraphics[width=0.5\columnwidth]{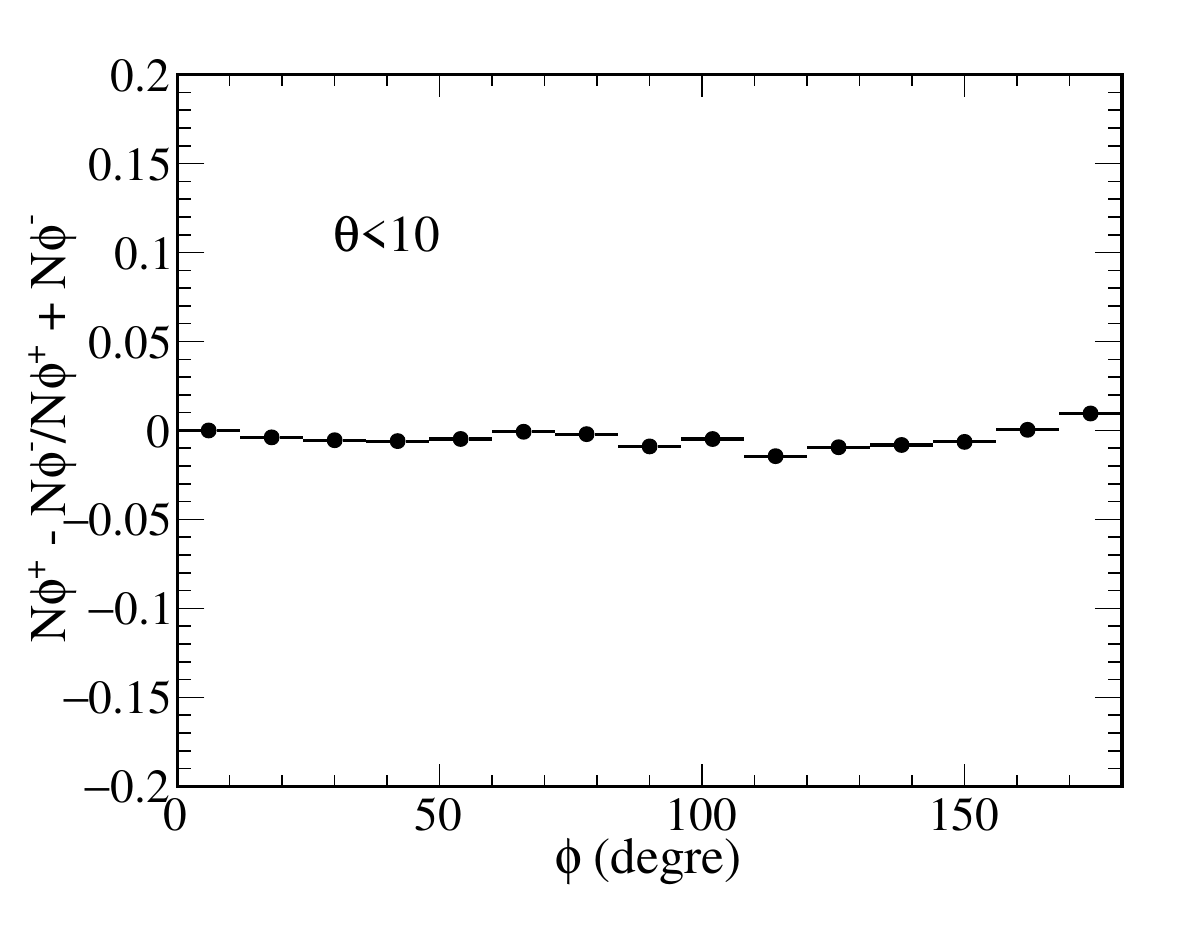}
\put(-100,50){Preliminary}\\
\caption{Experimental azimuthal counting asymmetry for $30^{\circ}<\theta<45^{\circ}$  left panel, for $\theta<10^{\circ}$ right panel.\label{genexp}}
\end{figure}

During the analysis of data collected by the EEE telescope located in the Department of Physics of the University of Genoa an asymmetry of the order of 10\%   on the counting rate has been observed between the muons N$_{\phi^+}$, coming from the valley side ($0^{\circ}<\phi\le 180^{\circ}$, namely the side with respect to telescope position in figure \ref{gen}) and the ones N$_{\phi^-}$, from the hill side ($-180^{\circ}<\phi\le 0^{\circ}$, namely the left side with respect to telescope in figure \ref{gen}). The asymmetry  appears more clearly in  the azimuthal angle distribution at polar angle $30^{\circ}<\theta< 45^{\circ}$, as reported in figure \ref{genexp}, left panel. This asymmetry seems to be due to the absorption by the asymmetric structure of the building. In fact for the polar angle range 30-45 degree  from the hill side the muons  should pass through 5 floors, the walls of the rooms and the roof, while in the opposite direction   ($-180^{\circ}<\phi\le 0^{\circ}$) the amount of absorber material is much less.
The role of the building  structure   is confirmed by the azimuthal asymmetry disappearing for small polar angles, where approximately the same absorber material shields the telescope. In fact for polar angles lower than 10 degree the asymmetry is flat around zero as reported in figure \ref{genexp}, right panel.

\begin{figure}[h]
\centering
\includegraphics[width=0.6\columnwidth]{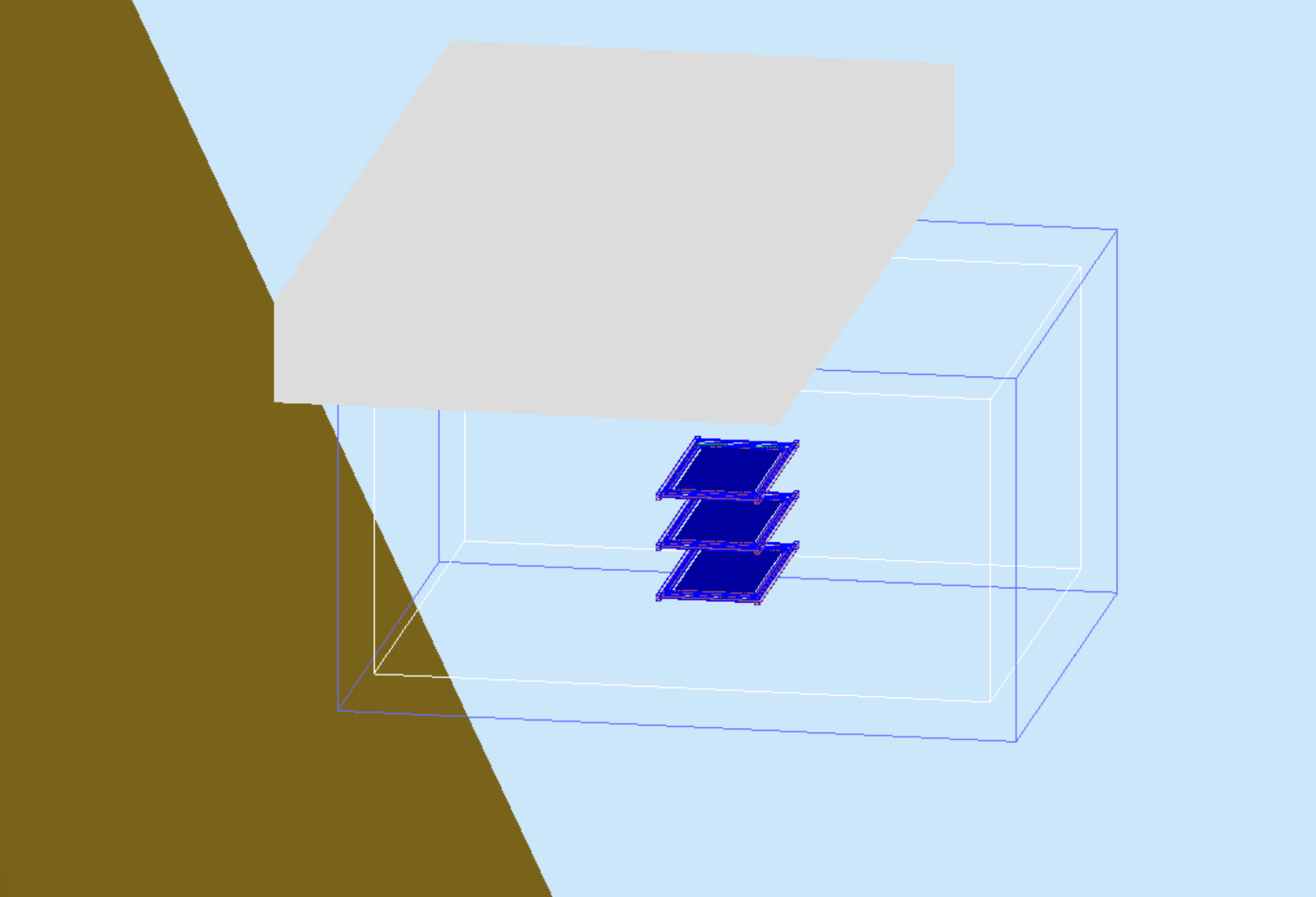}
\caption{Rendering 3D of geometry used to simulate the Genoa Physics Department building. \label{genrend}}
\end{figure}
In order to check the ability of the simulation to reproduce such effect, we generate a simulated data sample without any shielding  around the telescope and another  by parametrizing the building structure by placing a box of 70 cm of concrete (equivalent to 5 floors plus the roof) in the direction $-180^{\circ}<\phi\le 0^{\circ}$ above the room hosting the telescope, as shown in figure \ref{genrend}.

\begin{figure}[h]
\centering
\includegraphics[width=0.5\columnwidth]{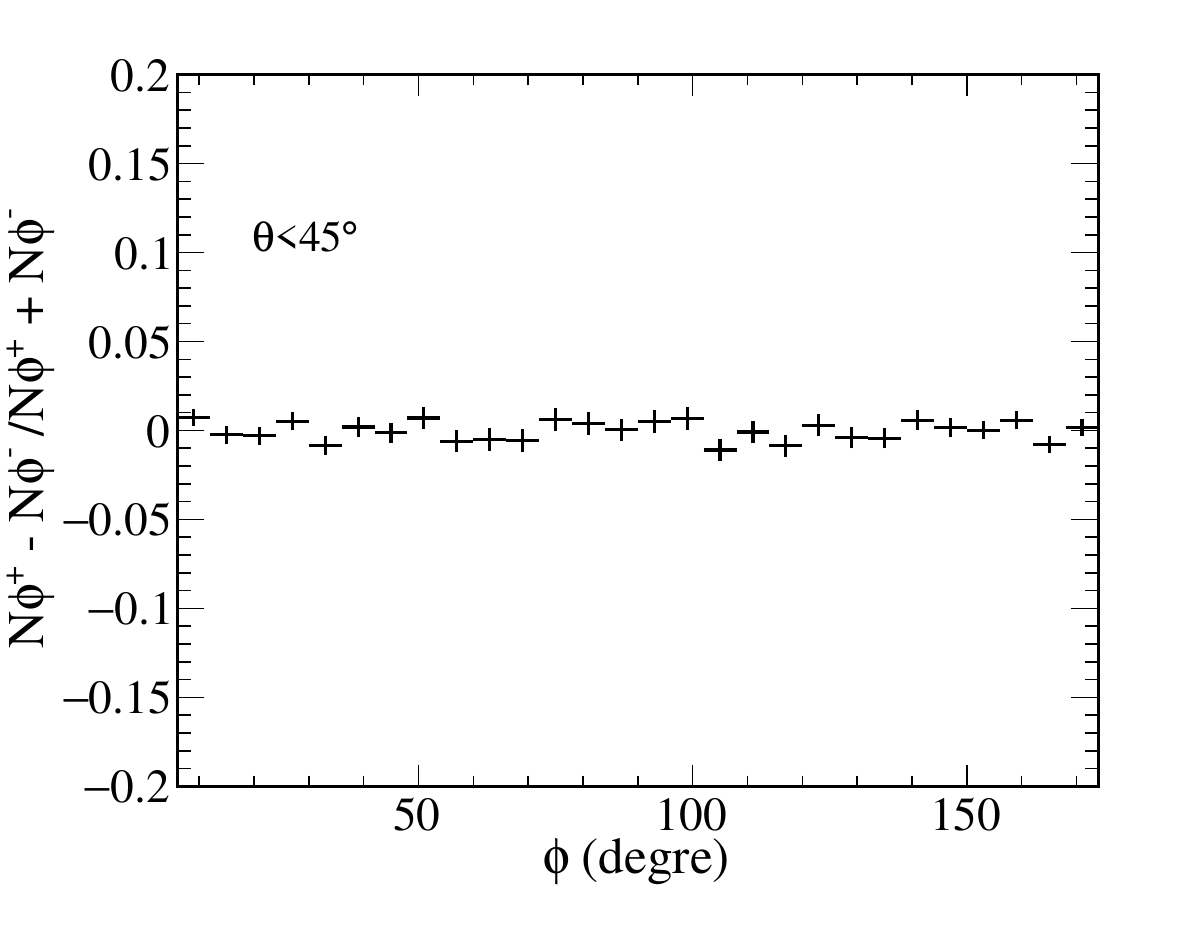}
\put(-100,40){Preliminary}
\includegraphics[width=0.5\columnwidth]{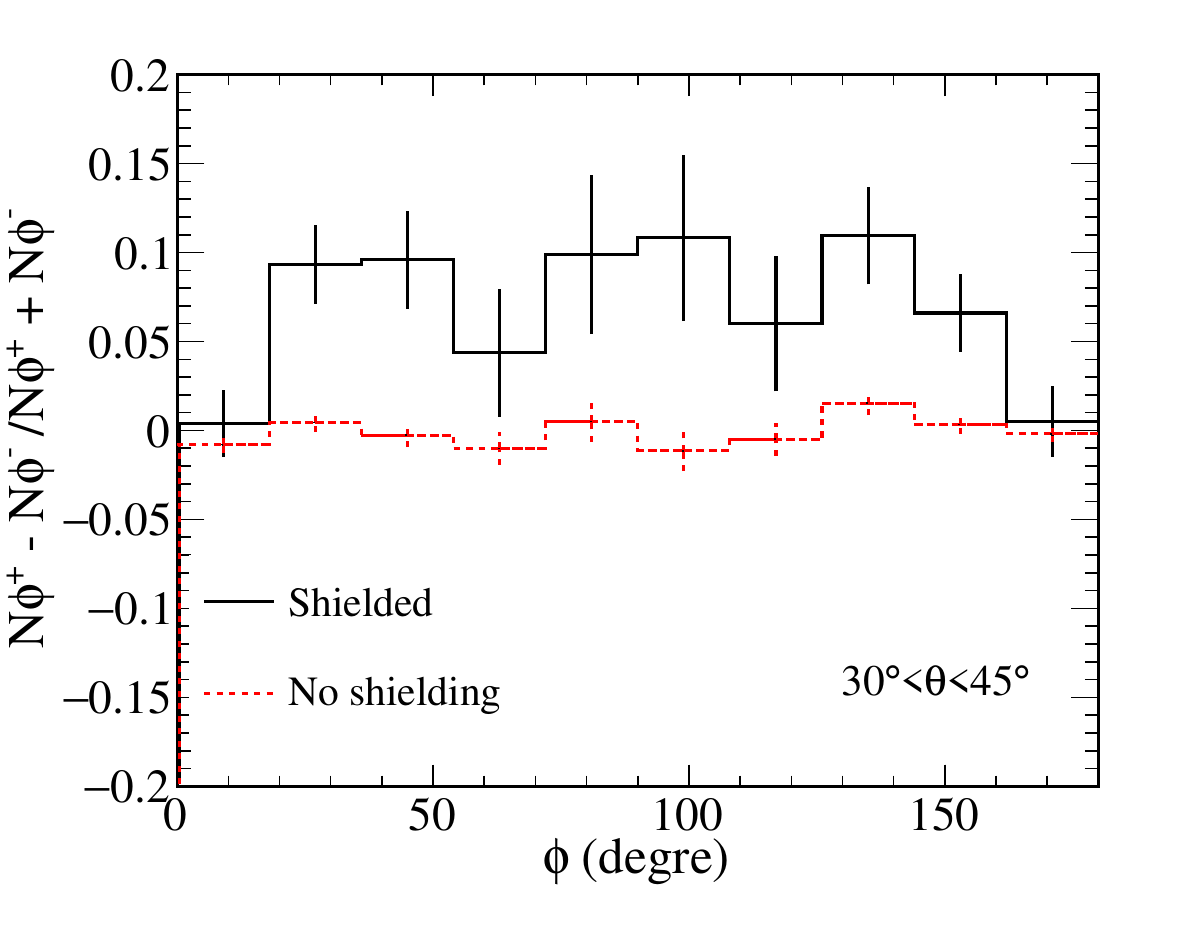}
\put(-130,50){Preliminary}
\caption{Left panel: azimuthal counting asymmetry for the simulated data with telescope without any shielding for $\theta<45^{\circ}$. Right panel: simulated counting asymmetry with a parametrized asymmetric building structure as in Fig. \ref{genrend} solid line, without any shielding dashed line  for $30^{\circ}<\theta<45^{\circ}$.\label{gensim}}
\end{figure}

The same procedure previously applied in  the analysis of the experimental data has been used to extract the asymmetry for both simulated data samples, for $30^{\circ}<\theta<45^{\circ}$ polar angles.
No asymmetry effect on the azimuthal distribution of the telescope without any shielding appeared, as reported in Fig.~\ref{gensim}, left panel. This proves the simulation does not produce any artificial asymmetry in the investigated distribution. 
Whereas, the distribution obtained by analysing a simulated data coming from the simplified structure parametrization of Department of Physics in Genoa  shows an asymmetry of the order of 10\% as the experimental one (see right panel of Figure \ref{gensim}). Of course, in this attempt just a crude parametrization of the  building structure was used, and this explains the slight difference in shape of the experimental and simulated distributions. Nevertheless, such a qualitative study proves the  simulation capability of reproducing realistic experimental conditions of data taking. 
    
\subsection{Detector resolution estimation}

\begin{figure}[h]
\centering
\includegraphics[width=0.6\columnwidth]{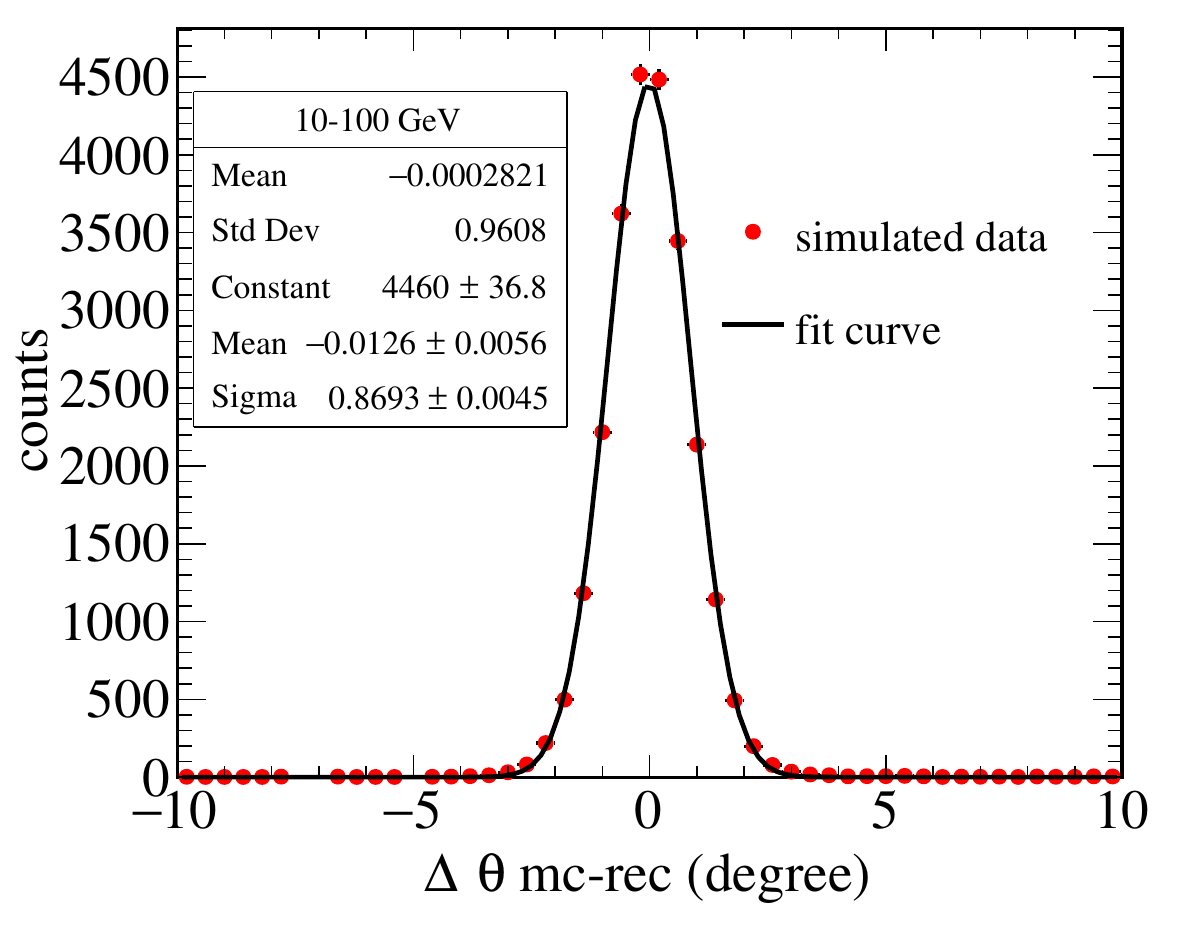}
\put(-80,40){\small Preliminary}
\caption{Difference between the generated and reconstructed polar angle of muons with energy of 10-100 GeV. Full circles simulated data, full line result of a Gaussian fit.} \label{ris}
\end{figure}

The resolution on the muon polar angle and on the hit position on the middle chamber  of the detectors were also estimated by analysing a simulated data sample generated with an EEE telescope in a space containing just air and by using only high-energy muons (higher than 10 GeV). Only high energy muons were used to perform this estimation to make  negligible the effects of multiple scattering from the air medium on the particle direction.
As an estimator of the resolution the standard deviation of the distributions obtained as the difference of generated and reconstructed polar angles was used, as reported in figure \ref{ris}.

%
%

A polar angle resolution better than 1 degree (standard deviation of 0.96 degree from simulated data, 0.87 degree from a Gaussian fit see Fig. \ref{ris}) was found. A similar study on position resolution resulted in a  spatial resolution $\sigma_{\Delta X}=$~1.64~cm  and $\sigma_{\Delta Y}=$~1.07~cm. These results are quite similar to the experimental resolution estimation reported in Ref. \cite{degruttola}, found to be  $\sigma_{\Delta X}= 1.47 \pm 0.23$~cm and  $\sigma_{\Delta Y}= 0.92 \pm 0.02$~cm for the X and Y position, respectively, and this
result proves once again the potentiality of this tool for a detailed model of this detector.

\section{Conclusion}
A simulation tool  to describe the EEE experiment MRPC telescope \cite{Ref1,Ref2} based on the GEMC framework\cite{gemc} was presented here. The event generator is implemented by using an improved version of the Gaisser parametrization of the  cosmic muon flux as a function of muon energy and momentum\cite{egpar}. 
A procedure to estimate the telescope efficiency directly from data was presented, which proved its reliability  by comparing experimental and simulated data.
Moreover, this tool is able to describe the single telescope behaviour reproducing an important quantity such as the muon polar angle direction  with a precision of 10\% in the whole detector acceptance. 
The behaviour of a telescope working in a building with a singular structure lying on the side of a hill was reproduced, showing the potentiality of the simulation tool.
The estimation of the detector resolution with the simulation has been performed showing a good agreement with the experimental result reported in Ref. \cite{degruttola}.

%
%
%
%
%
%


\end{document}